\documentclass{appolb}
\usepackage{graphicx,epsfig,amssymb}

\def   \lsth     {littlest Higgs}

\newcommand\iden{\leavevmode\hbox{\small1\normalsize\kern-.33em1}}
\def \nn {\nonumber}
\def\w{{\rm w}}

\def\gmm{{\gamma_{\mu}}}

\def\gff{{\gamma_{5}}}

\def\gvi{{g_{V_i}}}
\def\gai{{g_{A_i}}}

\def\mn{{g_{\mu\nu}}}

\def\boss{\hskip2mm }

\begin{document}
\title{The neutral heavy
scalar productions associated with $Z_L$ in the littlest Higgs model
at ILC and CLIC
}
\author{A. \c{C}A\~{G}IL$^*$ and M. T. ZEYREK$^\dagger$
\address{Physics Department, Middle East
Technical University,\\06531 Ankara, Turkey\\
$^*$ayse.cagil@cern.ch\\
$^\dagger$zeyrek@metu.edu.tr}
}
\maketitle
\begin{abstract}
In this work, the production processes of heavy neutral scalar and
pseudo scalar associated with standard model gauge boson $Z_L$ at
future $e^{+}e^{-}$ colliders (ILC and CLIC) are examined. The total
and differential cross sections are calculated for the processes in
the context of the littlest Higgs model. Also dependence of
production processes to littlest Higgs model parameters in the range
of compatibility with electroweak precision measurements and decays to lepton flavor violating final states 
are analyzed. We have found that both heavy scalar and pseudoscalar will be produced in $e^+e^-$ colliders. Also the depending on the model parameters, the neutral heavy scalar can be reconstructed or lepton flavor violating signals can be observed.

\end{abstract}
  
\section{Introduction}
Standard model (SM) is an effective theory with a cut off scale
around electroweak symmetry breaking (EWSB) scale. However in SM,
Higgs scalar, giving mass to fermions and gauge bosons gets loop
corrections to its mass up to cut off scale, which is called the
hierarchy problem. The little Higgs
models\cite{lh1,lhmodels1,lhmodels2,lhmodels3} are introduced to
solve the hierarchy problem among the alternative solutions such as
supersymmetry, extra dimensions and dynamical symmetry breaking
models. The little Higgs models propose a solution by enlarging the
symmetry group of the SM. The constraints on little Higgs models are
studied\cite{perelstein2ew,B1rizzo,Bdawson,Bkilian,Bdias,B2csaki,hakem1},
and the phenomenology of the little Higgs models are
reviewed\cite{perelstein1,thanrev,schmalzrev1}. The little Higgs
models are also expected to give significant signatures in future
high energy colliders and studied
\cite{atlaswork,atlaswork2,LHCreuter,A3,cagil1,cagil2}.

In the littlest Higgs model\cite{lh1} as a result of enlarged
symmetry group, there appears new vector gauge bosons and also a new
heavy scalar triplet. The appearance of new scalars in the littlest
Higgs model result in lepton flavor violation when a $5D$ operator
is implemented in the Yukawa
lagrangian\cite{thanlept1,gaurlept1,cinlept_L2yue,gaurlept2}.

In this work we examined the production of neutral scalar ($\phi^0$)
and pseudo scalar ($\phi^P$) associated with $Z_L$ boson in the
littlest Higgs model at $e^{+}e^{-}$ colliders, namely,
International Linear Collider (ILC) \cite{ILC} and Compact Linear
Collider (CLIC) \cite{CLIC}. To analyze the production
rates, firstly the most promising channel $e^+e^-\rightarrow Z_L
\phi^0$ is analyzed. Since the final signals of $\phi^0$ and
$\phi^P$ is same, to analyze the behavior of the final states for
the energies ($\sqrt{s}>1TeV$), the higher order production
processes; $e^+e^-\rightarrow Z_L \phi^0\phi^0$, $e^+e^-\rightarrow
Z_L \phi^P\phi^P$ and $e^+e^-\rightarrow Z_L \phi^0\phi^P$  are also
examined. Since the process $e^+e^-\rightarrow Z_L \phi^P$ is not
allowed in the littlest Higgs model, the latter two processes
involving $\phi^P$ are also important for the $\phi^P$ production. 
Finally, lepton flavor violating signals of neutral scalars as \emph{``$Z_L+$missing energy''} which characterizes the new 
neutral scalar and pseudoscalar to be littlest Higgs are analyzed\cite{thanlept1}.

In this paper, we present the relevant formulas and calculations in
section $2$. In section $3$, the results and discussions are
presented.

\section{Theoretical Framework}

In the littlest Higgs model global symmetry $SU(5)$ is broken
spontaneously to $SO(5)$ at an energy scale $f\sim 1TeV$ leaving
$14$ Nambu  Goldstone bosons(NGB) corresponding to broken symmetries.
In the model $SU(5)$ contains the gauged subgroup $[SU(2)_1\otimes
U(1)_1]\otimes[SU(2)_2\otimes U(1)_2]$. As a consequence symmetry
breaking, gauge bosons gain mass by eating the four of the NGBs. The
mixing angles between the $SU(2)$ subgroups and between the $U(1)$
subgroups are defined respectively as:
\begin{equation}\label{ssp}
    s\equiv\sin\theta=\frac{g_2}{\sqrt{g_{1}^2 + g_{2}^2 }}~~,~~~~ s^\prime \equiv\sin\theta'=\frac{g'_2}{\sqrt{g_{1}^{\prime 2} + g_{2}^{\prime
    2}
    }}~~,
\end{equation}
where $g_i$ and $g'_i$ are the gauge couplings of $SU(2)_i$ and
$U(1)_i$ subgroups respectively.
By EWSB vector bosons get extra mixings due to vacuum expectation
values of $h$ doublet and $\phi$ triplet resulting the final masses
to the order of $\frac{v^2}{f^2}$ such as\cite{thanrev}:
\begin{eqnarray}\label{massesvectors}
    M_{A_L}^2 &=& 0 ,\nonumber \\
    M_{Z_L}^2 &=& m_z^2
    \left[ 1 - \frac{v^2}{f^2} \left( \frac{1}{6}
    + \frac{1}{4} (c^2-s^2)^2
    + \frac{5}{4} (c^{\prime 2}-s^{\prime 2})^2 \right)
    + 8 \frac{v^{\prime 2}}{v^2} \right],
    \nonumber \\
    M_{A_H}^2 &=&
    \frac{f^2 g^{\prime 2}}{20 s^{\prime 2} c^{\prime 2}}
    - \frac{1}{4} g^{\prime 2} v^2 + g^2 v^2 \frac{x_H}{4s^2c^2}\\
\nn& =& m_z^2 s_{\w}^2 \left(
    \frac{ f^2 }{5 s^{\prime 2} c^{\prime 2}v^2}
    - 1 + \frac{x_H c_{\w}^2}{4s^2c^2  s_{\w}^2} \right),
    \nonumber \\
    M_{Z_H}^2 &=& \frac{f^2g^2}{4s^2c^2}
    - \frac{1}{4} g^2 v^2
    - g^{\prime 2} v^2 \frac{x_H}{4s^{\prime 2}c^{\prime 2}}\\
   \nn & =& m_w^2 \left( \frac{f^2}{s^2c^2 v^2}
    - 1 -  \frac{x_H s_{\w}^2}{s^{\prime 2}c^{\prime 2}c_{\w}^2}\right) ,
\end{eqnarray}
where   $m_z\equiv {gv}/(2c_{\w})$ and $x_H = \frac{5}{2} g
g^{\prime}
    \frac{scs^{\prime}c^{\prime} (c^2s^{\prime 2} + s^2c^{\prime 2})}
    {(5g^2 s^{\prime 2} c^{\prime 2} - g^{\prime 2} s^2 c^2)}$ and
$s_{\w}$ and $c_{\w}$ are the usual weak mixing angles. The parameters
$v$ and $v'$ are the vacuum expectation values of scalar doublet and
triplet given as\cite{thanrev};
\begin{equation}\label{vandvprimelimits}
\langle h^0 \rangle = v/\sqrt{2}~~,~~~~\langle i \phi^0 \rangle =
v^{\prime}\leq \frac{v^2}{4f},
\end{equation}
bounded by electroweak precision data, where $v=246 GeV$. Also
diagonalizing the mass matrix for scalars the physical states are
found to be the SM Higgs scalar $H$, the neutral scalar $\phi^0$,
the neutral pseudo scalar $\phi^P$, and the charged scalars $\phi^+$
and $\phi^{++}$. The masses of the heavy scalars are degenerate, and
in terms of Higgs mass expressed as\cite{thanrev}:
\begin{eqnarray}
    M_\phi =\frac{\sqrt{2} f}{v\sqrt{1-(\frac{4 v'
    f}{v^2})^2}}M_H.
\end{eqnarray}

The scalar fermion interactions in the model are written in Yukawa
lagrangian preserving gauge symmetries of the model for SM leptons
and quarks, including the third generation having an extra singlet,
the $T$ quark. The fermions in the littlest Higgs model can be
charged under both $U(1)_1$ and $U(1)_2$
subgroups\cite{B2csaki,thanrev}. Also for light fermions, a lepton
number violating coupling can be implemented in Yukawa
lagrangian\cite{thanlept1,gaurlept1} which results in lepton flavor
violation by unit two, such as:
\begin{equation}\label{lepviol1}
    {\cal L}_{LFV} = iY_{ij} L_i^T \ \phi \, C^{-1} L_j + {\rm h.c.},
\end{equation}
where $L_i$ are the lepton doublets $\left(
                                       \begin{array}{cc}
                                         l &\nu_l \\
                                       \end{array}
                                     \right)$,
and $Y_{ij}$ are the elements of the mixing matrix with $Y_{ii}=Y$ and $Y_{ij(i\neq
j)}=Y'$ . The values of Yukawa couplings $Y$ and $Y'$ are restricted
by the current constraints on the neutrino
masses\cite{neutrinomass}, given as; $M_{ij}=Y_{ij}v'\simeq
10^{-10}GeV$\cite{thanlept1}. Since the vacuum expectation value
$v'$ has only an upper bound(Eq. \ref{vandvprimelimits}), $Y_{ij}$ can be taken up to
order of unity without making $v'$ unnaturally small.

The parameters  $f$; the symmetry breaking scale , and $s,s'$; the
mixing angles of the {\lsth} model are not restricted by the model.
These parameters are constrained by observables of electroweak
precision data and the direct search for a heavy gauge bosons at
Tevatron\cite{perelstein2ew,B1rizzo,Bdawson,Bkilian,Bdias,B2csaki}.
In the case when fermions are charged under both $U(1)$ groups, the
allowed parameter space is listed as follows. For the values of the
symmetry breaking scale $1TeV \leq f \leq 2 TeV$, mixing angles are
in the range $0.75\leq s \leq 0.99$ and $0.6\leq s' \leq 0.75 $, for
$ 2TeV \leq f \leq 3TeV$ they have acceptable values in the range
$0.6\leq s \leq 0.99$ and $0.6\leq s' \leq 0.8$, for $3TeV \leq f
\leq 4 TeV$ they are in the range $0.4\leq s \leq 0.99$ and $0.6\leq
s' \leq 0.85$, and for the higher values of the symmetry breaking
scale, i.e. $f\geq 4TeV$, the mixing angles are less restricted and
they are in the range $0.15\leq s \leq 0.99$ and $0.4\leq s' \leq
0.9$ \cite{B2csaki}.


\begin{table}[htb]
\begin{center}
\caption{The vector and axial vector couplings of $e^{+}e^{-}$ with
vector bosons. Feynman rules for $e^{+}e^{-} V_i$ vertices are given
as $i \gmm (\gvi + \gai \gff)$ .\label{gVgA}} 
\begin{tabular}{|c||c|c|c|}
  \hline
   $i$& vertices &$\gvi$  &$\gai$    \\
  \hline\hline 1&$e^{+}e^{-} Z_L$ &
    $\frac{-g}{2c_{\w}} \left\{ -\frac{1}{2} + 2 s^2_{\w}- \frac{v^2}{f^2} \left[ \frac{-c c_{\w} x_Z^{W^{\prime}} }{2s}
    \right. \right.$ &
    $\frac{-g}{2c_{\w}} \left\{ \frac{1}{2}
    - \frac{v^2}{f^2} \left[ \frac{c c_{\w} x_Z^{W^{\prime}} }{2s}
    \right. \right.$ \\
& &  $\left. \left.
    + \frac{s_{\w} x_Z^{B^{\prime}}}{s^{\prime}c^{\prime}}
    \left( 2y_e - \frac{9}{5} + \frac{3}{2} c^{\prime 2}
    \right) \right] \right\}$&
    $\left. \left.
    + \frac{s_{\w} x_Z^{B^{\prime}}}{s^{\prime}c^{\prime}}
    \left( -\frac{1}{5} + \frac{1}{2} c^{\prime 2} \right)
    \right] \right\}$ \\
  \hline 2& $e^{+}e^{-} Z_H$ &$-gc/4s$ & $gc/4s$      \\
  \hline 3&$e^{+}e^{-} A_H$  &
    $\frac{g^{\prime}}{2s^{\prime}c^{\prime}}
    \left( 2y_e - \frac{9}{5} + \frac{3}{2} c^{\prime 2} \right)$ &
    $\frac{g^{\prime}}{2s^{\prime}c^{\prime}}
    \left( -\frac{1}{5} + \frac{1}{2} c^{\prime 2} \right)$    \\
  \hline4& $e^{+}e^{-} A_L$&$-e$&0     \\
  \hline
\end{tabular}

\end{center}

\end{table}

In the model, the couplings of vector bosons to fermions are written
as $i \gmm (\gvi + \gai \gff)$ where $i=1,2,3,4$ corresponds to
$Z_L$, $Z_H$, $A_H$ and $A_L$ respectively. The couplings of gauge
vector to $e^{+}e^{-}$ pairs are given in table \ref{gVgA}, where
$y_e=\frac{3}{5}$, $e=\sqrt{ 4 \pi \alpha}$, $x_Z^{W^{\prime}} =
-\frac{1}{2c_{\w}} sc(c^2-s^2)$ and $x_Z^{B^{\prime}} =
-\frac{5}{2s_{\w}} s^{\prime}c^{\prime}
    (c^{\prime 2}-s^{\prime 2})$. The
total decay widths of SM vector bosons also get corrections of the
order $\frac{v^2}{f^2}$, since the decay widths of vectors to
fermion couples are written as; $\Gamma(V_i\rightarrow
f\bar{f})=\frac{N}{24\pi}(g^2_V+g^2_A)M_{V_i}$ where $N=3$ for
quarks, and $N=1$ for leptons. The total decay widths of the new
vector bosons are given as \cite{A3}:
\begin{eqnarray}
 \nn\Gamma_{A_H}&\approx&  \frac{g'^2 M_{A_H}(21-70 s'^2 +59
s'^4)}{48 \pi s'^2 (1-s'^2) },\\
 \Gamma_{Z_{H}}&\approx& \frac{g^2
(193 - 388 s^2 + 196 s^4)}{768 \pi s^2 (1-s^2)}M_{Z_H}.
\end{eqnarray}

The final decays and also the decay widths of $\phi^0$ and $\phi^P$
are studied in detail in Ref.\cite{thanlept1}, and they are strongly dependent on the VEV of the 
scalar triplet; $v'$. For $v'\gtrsim 1GeV$, the decay modes of
$\phi^0$ include decays in to quark pairs; $t\bar{t}$,
$b\bar{b}$ and $t\bar{T}$+$T\bar{t}$, and also decays into SM pairs;
$Z_L Z_L$ and $H H$. In this case the decays of $\phi^P$ are similar to $\phi^0$
as the decays in to quark pairs; $t\bar{t}$,
$b\bar{b}$ and $t\bar{T}$+$T\bar{t}$, and to SM $Z_L H$ couples different from $\phi^0$. For $v'\sim10^{-10}GeV$,
the non leptonic decays are suppressed by
a factor of $\frac{v'}{v}$ for both $\phi^0$ and $\phi^P$, and the final states contain only lepton flavor violating decays to $\nu_i \nu_j +
\bar{\nu_i}\bar{\nu_j}$. In this work, we analyze the cases
$v'\sim 1GeV~~(Y<<1)$ and $v'=10^{-10}GeV~~( Y\sim 1)$. The decay
widths of scalars in these cases can be written as\cite{thanlept1}:
\begin{eqnarray}\label{dwp2}
\nn\Gamma_{\phi(v'\sim1)}&\simeq&\frac{N_c M_\phi }{32 \pi f^2}(M^2_b+M^2_t)+\frac{v^{\prime 2}M^3_\phi }{2\pi v^4},\\
\Gamma_{\phi(v'\sim 10^{-10})}&\simeq& \Gamma_{\phi (LFV)}=
\frac{| Y|^2}{8\pi
 } M_\phi .
\end{eqnarray}

\begin{figure}[htb]
\begin{center}

\includegraphics[width=5cm]{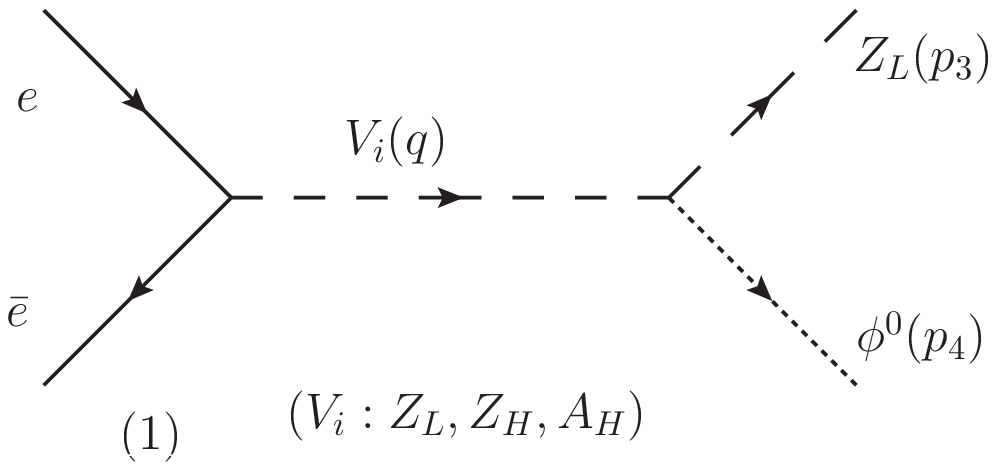}

\qquad\allowbreak
\end{center}
\caption{Feynman diagrams contributing to  $e^{+}e^{-}\rightarrow
Z_L \phi^0$ in {\lsth} model.} \label{fd3}
\end{figure}

The properties of new neutral scalar $\phi^0$, its couplings to SM
and new neutral vector bosons can be examined in single production
of $\phi^0$ associated with $Z_L$ events. The couplings of $\phi^0$
to $Z_L$ and vectors are in the form $i g_{\mu\nu} B_i$, where
$i=1,2,3$ corresponds to $Z_L,Z_H,A_H$ respectively and given in
table \ref{pvvcouplings}, where $s_0 \simeq  2 \sqrt{2}
\frac{v^{\prime}}{v}$. The Feynman diagrams contributing this
process are given in figure \ref{fd3}.
%

%
\begin{table}[htb]

\begin{center}

\caption{The Feynman rules for $\phi^0 V_i V_j$ vertices.\label{pvvcouplings}}

{\begin{tabular}{|c||c|c||c||c|c|}
  \hline
  i/j&vertices & $i g_{\mu\nu} B_{ij}$ \\
\hline\hline  1/1 &$\phi^0 Z_L Z_L$& $-\frac{i}{2}
\frac{g^2}{c^2_\w} ( v s_0 - 4 \sqrt{2} v^{\prime} )
    g_{\mu\nu}$\\
    \hline
 2/2& $\phi^0 Z_H Z_H$ & $\frac{i}{2} g^2 \left( v s_0
    + \frac{(c^2-s^2)^2}{s^2c^2} \sqrt{2} v^{\prime} \right) g_{\mu\nu}$ \\
\hline  1/2&$\phi^0 Z_L Z_H$ & $\frac{i}{2} \frac{g^2}{c_\w}
\frac{(c^2-s^2)}{2sc}( v s_0 - 4 \sqrt{2} v^{\prime} ) g_{\mu\nu}$
\\
\hline 2/3&$\phi^0 Z_H A_H$ & $\frac{i}{4} g g^{\prime}
\frac{1}{scs^{\prime}c^{\prime}} \left(
    (c^2s^{\prime 2} + s^2c^{\prime 2}) v s_0 \right.$\\
\hline  1/3&$\phi^0 Z_L A_H$ &$\frac{i}{2} \frac{gg^{\prime}}{c_\w}
    \frac{(c^{\prime 2}-s^{\prime 2})}{2s^{\prime}c^{\prime}}
    ( v s_0 - 4 \sqrt{2} v^{\prime} ) g_{\mu\nu}$\\
    \hline
    3/3&$\phi^0 A_H A_H$ & $\frac{i}{2} g^{\prime 2} \left( v s_0
    + \frac{(c^{\prime 2}-s^{\prime 2})^2}{s^{\prime 2}c^{\prime 2}}
    \sqrt{2} v^{\prime} \right) g_{\mu\nu}$\\
\hline
\end{tabular}}
\end{center}

\end{table}

\begin{figure}[h]
\begin{center}
\includegraphics[width=5cm]{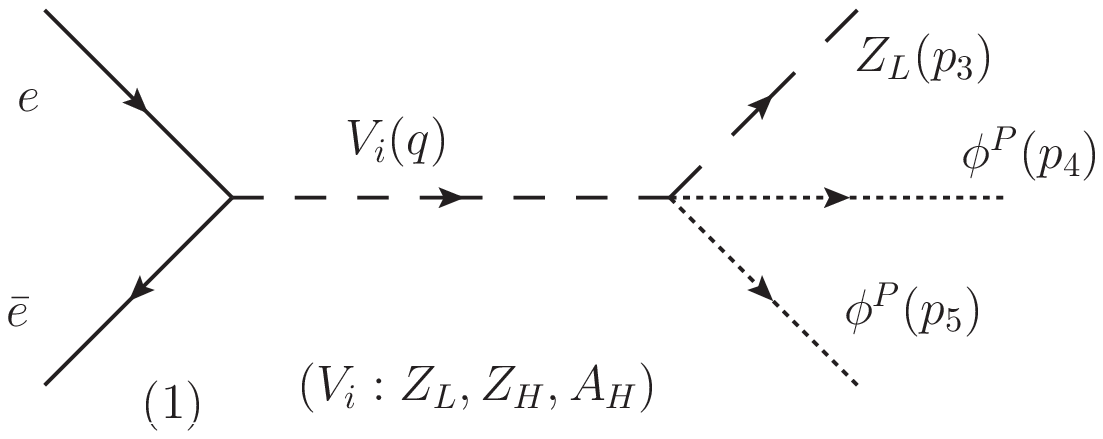}\boss\boss\includegraphics[width=5cm]{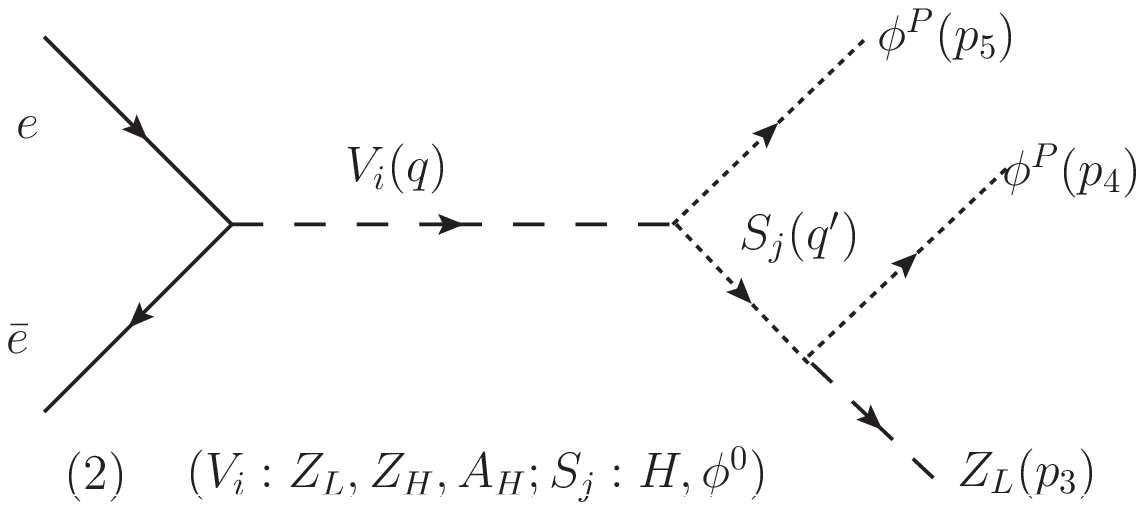}
\vskip2mm
\end{center}\caption{Feynman diagrams
contributing to $e^{+}e^{-}\rightarrow Z_L \phi^P \phi^P$ in {\lsth}
model.} \label{fd8}
\end{figure}

\begin{figure}[h]
\begin{center}
\includegraphics[width=5cm]{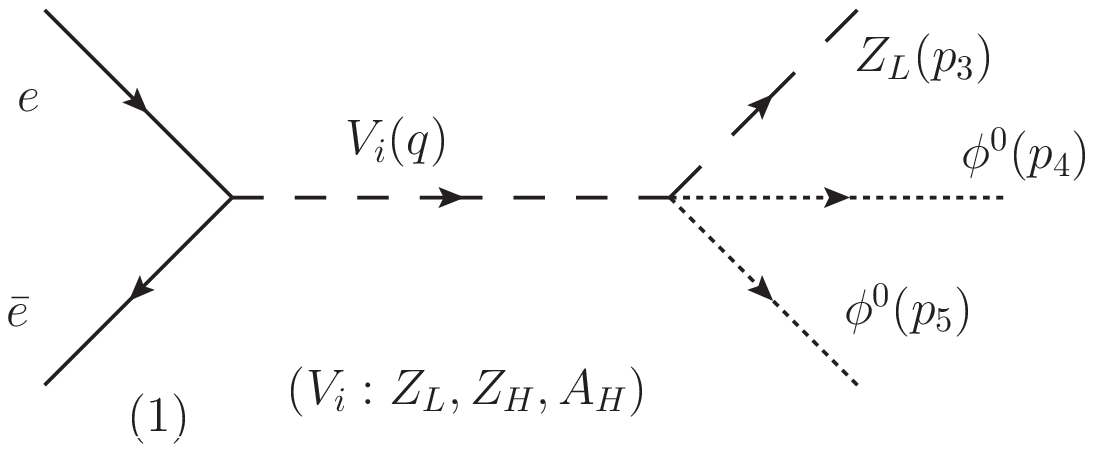}\boss\boss\includegraphics[width=5cm]{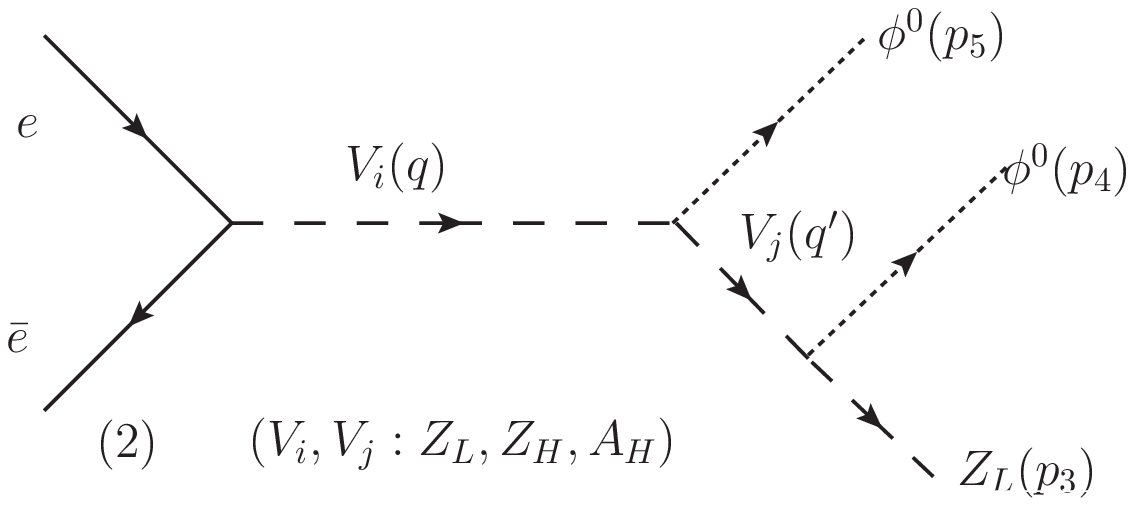}\boss
\qquad\allowbreak \vskip5mm
\includegraphics[width=5cm]{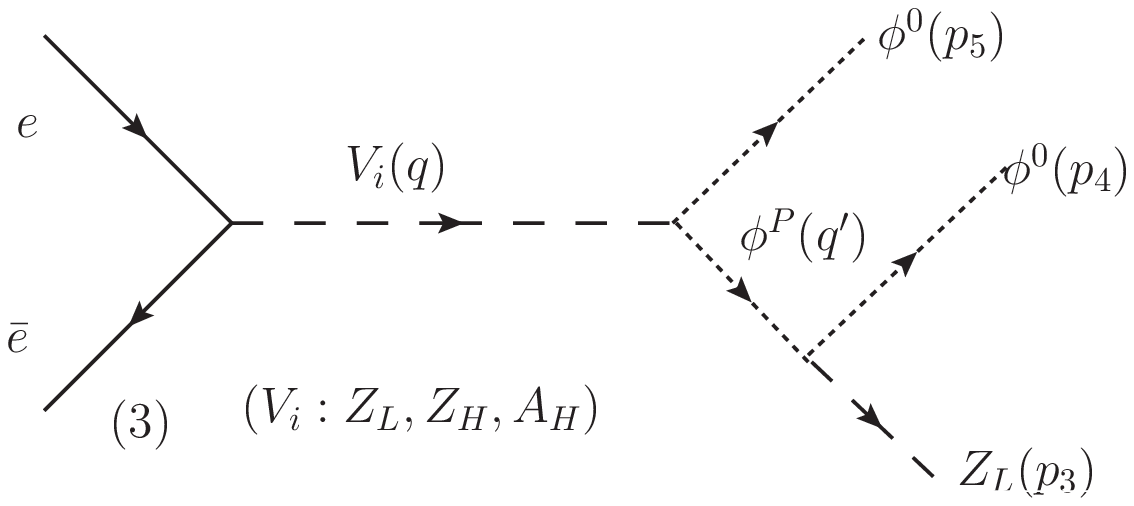}
\end{center}\caption{Feynman diagrams
contributing to $e^{+}e^{-}\rightarrow Z_L \phi^0 \phi^0$ in {\lsth}
model.} \label{fd6}
\end{figure}

\begin{figure}[htb]
\begin{center}
\includegraphics[width=5cm]{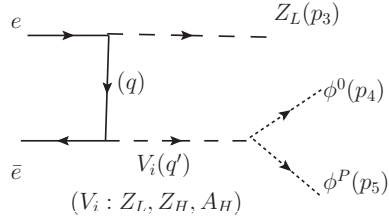}

\qquad\allowbreak
\end{center}
\caption{Feynman diagrams contributing to $e^{+}e^{-}\rightarrow Z_L
\phi^0 \phi^P$ in {\lsth} model.} \label{fd3b}
\end{figure}

The pair productions of neutral heavy scalar and pseudo scalar
associated with $Z_L$ via $e^{+}e^{-}\rightarrow Z_L\phi^0 \phi^0$,
$e^{+}e^{-}\rightarrow Z_L\phi^P \phi^P$ and $e^{+}e^{-}\rightarrow
Z_L\phi^0 \phi^P$ are also examined in this work. The Feynman rules
for scalar-vector couplings are given in table \ref{pvvcouplings},
the Feynman rules for four point scalar(pseudoscalar)-vector
couplings are given in table \ref{PPVVcouplings} and the Feynman
rules for pseudoscalar-vector-scalar couplings are  given in table
\ref{PpVScouplings}, where $s_P = \frac{2 \sqrt{2} v^{\prime}}
    {\sqrt{v^2 + 8 v^{\prime 2}}}
    \simeq 2 \sqrt{2} \frac{v^{\prime}}{v}$. The Feynman diagrams for
    the processes $e^{+}e^{-}\rightarrow Z_L\phi^0 \phi^0$,
$e^{+}e^{-}\rightarrow Z_L\phi^P \phi^P$ and $e^{+}e^{-}\rightarrow
Z_L\phi^0 \phi^P$ are presented in figures \ref{fd8}, \ref{fd6} and
\ref{fd3b} respectively.
\begin{table}[htb]

\begin{center}

\caption{The Feynman rules for four point interaction vertices between
scalars and vectors. Their couplings are given in the form $i C_{ij}
\mn $ and $i C^P_{ij} \mn $ respectively for $\phi^0 \phi^0 V_i V_j$
and $\phi^P \phi^P V_i V_j$. \label{PPVVcouplings}}
{
\begin{tabular}{|c||c|c||c|c|}
  \hline
  i/j&vertices & $i C_{ij} \mn $& vertices & $i C^P_{ij} \mn $ \\
\hline\hline  1/1 &$\phi^0 \phi^0 Z_L Z_L$&$2 i \frac{g^2}{c^2_\w}
g_{\mu\nu}$
& $\phi^P \phi^P Z_L Z_L$ &  $2 i \frac{g^2}{c^2_\w} g_{\mu\nu}$\\
\hline  1/2&$\phi^0 \phi^0 Z_L Z_H$ &$-2 i \frac{g^2}{c_\w}
\frac{(c^2-s^2)}{2sc}
        g_{\mu\nu}$
&$\phi^P \phi^P Z_L Z_H$ & $-2 i \frac{g^2}{c_\w}
\frac{(c^2-s^2)}{2sc}
        g_{\mu\nu}$\\
\hline  1/3&$\phi^0 \phi^0 Z_L A_H$ &$-2 i \frac{gg^{\prime}}{c_\w}
    \frac{(c^{\prime 2}-s^{\prime 2})}{2s^{\prime}c^{\prime}}
        g_{\mu\nu}$
&$\phi^P \phi^P Z_L A_H$ & $-2 i \frac{gg^{\prime}}{c_\w}
    \frac{(c^{\prime 2}-s^{\prime 2})}{2s^{\prime}c^{\prime}}
        g_{\mu\nu}$\\
\hline
\end{tabular}
}
\end{center}
\end{table}

%
%
%

%
\begin{table}[tbh]

\begin{center}

\caption{The Feynman rules for $\phi^P V_i S_j$ vertices.\label{PpVScouplings}}{

\begin{tabular}{|c||c|c||c||c|c|}
  \hline
  i/j&vertices & $-iE^P_{ij} (p_j - p_{\phi^P})$ \\
\hline\hline  1/1 &$\phi^P H Z_L$& $\frac{1}{2} \frac{g}{c_\w}
\left( s_P - 2 s_0 \right)
    (p_{\phi^P}-p_H)_{\mu}$\\
    \hline 1/2 & $\phi^P \phi^0 Z_L$ &  $-\frac{g}{c_\w} (p_{\phi^P}-p_{\phi^0})_{\mu}$ \\
\hline  2/1&$\phi^P H Z_H$ &$-\frac{1}{2} g \frac{(c^2-s^2)}{2sc}
    (s_P - 2 s_0)(p_{\phi^P}-p_H)_{\mu}$\\
\hline 2/2&$\phi^P \phi^0 Z_H$ & $g\frac{(c^2-s^2)}{2sc} (p_{\phi^P}-p_{\phi^0})_{\mu}$ \\
\hline  3/1&$\phi^P H A_H$ &$-\frac{1}{2}
    g^{\prime} \frac{(c^{\prime 2}-s^{\prime 2})}{2s^{\prime}c^{\prime}}
    (s_P - 2 s_0) (p_{\phi^P}-p_H)_{\mu}$\\
\hline 3/2&$\phi^P \phi^0 A_H$ &$g^{\prime} \frac{(c^{\prime
2}-s^{\prime 2})}{2s^{\prime}c^{\prime}}
    (p_{\phi^P}-p_{\phi^0})_{\mu}$ \\
\hline
\end{tabular}
}

\end{center}

\end{table}
\clearpage

\section{Results and discussions}

In this section the results for the processes
$e^{+}e^{-}\rightarrow Z_L \phi^0$, $e^{+}e^{-}\rightarrow Z_L
\phi^0 \phi^0$, $e^{+}e^{-}\rightarrow Z_L \phi^P \phi^P$ and
$e^{+}e^{-}\rightarrow Z_L \phi^{0} \phi^{P}$ are presented. The numerical values of the input parameters are taken to be: 
the Higgs mass $M_H=120GeV$ and the masses of standard model bosons
$M_{Z_L}=91GeV$, $M_{W_L}=80GeV$, and the fine structure constant
$\alpha=1/137.036$, consistent with recent data\cite{pdg}. The numerical calculations of cross
sections of the production processes are performed by {\tt
CalcHep}\cite{calchep} generator after implementing necessary
vertices.

\begin{figure}[h]
\begin{center}
\includegraphics[width=7cm]{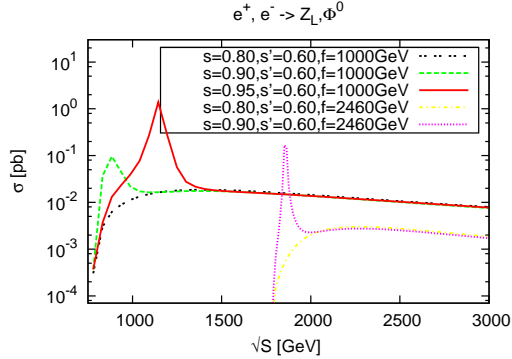}
\caption{Total cross section versus center of mass energy graphs of the process $e^{+}e^{-}\rightarrow Z_L \phi^0$ for some selected values of littlest Higgs model parameters when $v'=1GeV$. \label{f3x4}}
\end{center}
\end{figure}

For the examination of the production of heavy neutral scalar $\phi^0$ at linear colliders, 
the single production of $\phi^0$ associated with $Z_L$ is the most dominant channel. For this process, 
total cross section is plotted with respect to center of mass energy in figure \ref{f3x4} for different values 
of symmetry breaking scale $f$ and mixing angles $s$ and $s'$ allowed by recent constraints. In these calculations 
the VEV of the scalar triplet is taken to be $v'=1GeV$ allowed by the limit given in equation \ref{vandvprimelimits}. It is seen from figure \ref{f3x4} that, for symmetry breaking scale $f=1TeV$, and $s/s'=0.80/0.60$, the total cross section is at the order of 
$10^{-2}pb$ for $\sqrt{S}\sim 1000 - 3000 GeV$. For parameters $f=1TeV$ and $s/s'=0.90/0.60(0.95/0.60)$, the total cross section is at the order of 
$10^{-2}pb$ for $\sqrt{S}\geq 1200 (1500)GeV$. Also for these parameter sets, a resonance corresponding to heavy gauge boson
 $Z_H$ exist at $\sqrt{S}\sim900(1100)GeV$ increasing the total cross section up to $0.1(1)pb$. Unfortunately these 
resonances can have significance only if $\phi^0$ can be reconstructed. For $f=2.5TeV$, the cross section versus $\sqrt{S}$ graphs are also 
plotted in figure \ref{f3x4}, for mixing angles $s/s'=0.80/0.60$ and $s/s'=0.90/0.60$. In both set of parameters 
total cross section is about $5\times 10^{-3}pb$ for $\sqrt{S}\gtrsim 2TeV$. Also for $s/s'=0.90/0.60$, total cross section receives a peak up to $0.1pb$ about 
$\sqrt{S}\sim 1.8TeV$ corresponding to the resonance of $Z_H$. Finally the production of $\phi^0$ via $e^{+}e^{-}\rightarrow Z_L \phi^0$ process
is possible for low values of symmetry breaking scale $f=1TeV$, for both ILC($\sqrt{S}=1TeV$) and CLIC($\sqrt{S}=3TeV$). However for 
higher values of $f$, this channel is not promising.

For $v'\sim 1GeV$, the neutral scalar $\phi^0$ dominantly decays into quark pairs $t\bar{t}$ and $t\bar{T} + \bar{t} T$, with branching ratios of $0.8$ and $0.2$ respectively\cite{thanlept1}. Thus, the channel $e^+e^-\to Z_L t \bar{t}$ is promising for $\phi^0$ observation. In this channel, there will be more than thousands events which are observable as a contribution of $e^+e^-\to Z_L \phi^0$ process. Also in this channel, the SM background is at the order of $10^{-2}pb$ at $\sqrt{S}=1TeV$, and reduces to $10^{-4}pb$ for $\sqrt{S}\sim 2TeV$. So, for $\sqrt{S}\geq 1TeV$, the collider signal $Z_L t\bar{t}$ is dominated by the decays of neutral scalars produced via $e^+e^-\to Z_L \phi^0$ process. And also, by applying a cut on the energy of final state $t\bar{t}$ pair, i.e. $E_{t\bar{t}}\geq M_{\phi}$, will suppress the background contribution from SM. So in this channel, $\phi^0$ can be observed and reconstructed from $t\bar{t}$ jets for $\sqrt{S}\geq 1TeV$.

\begin{figure}[h]
\begin{center}
\includegraphics[width=7cm]{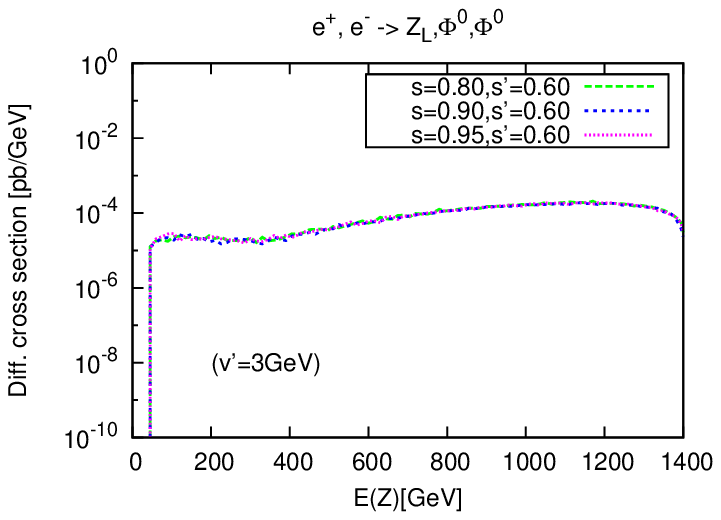}\caption{ Differential cross section versus energy of the $Z_L$ 
boson garphs of the process $e^{+}e^{-}\rightarrow Z_L
\phi^0 \phi^0$ for some selected values of mixing angles when $f=1TeV$ and $v'=3GeV$ at $\sqrt{S}=3TeV$.\label{fD1}}
\end{center}
\end{figure}

\begin{figure}[h]
\begin{center}
\includegraphics[width=7cm]{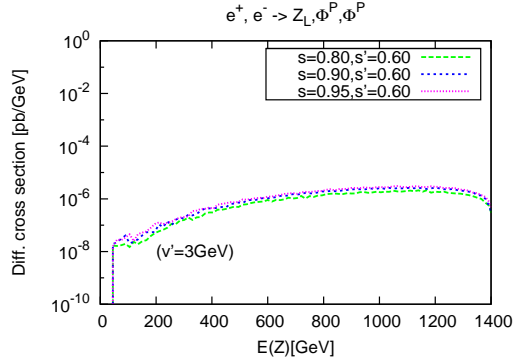}\caption{ Differential cross section versus energy of the $Z_L$ 
boson graphs of the process $e^{+}e^{-}\rightarrow Z_L
\phi^P \phi^P$ for some selected values of mixing angles when $f=1TeV$ and $v'=3GeV$ at $\sqrt{S}=3TeV$.\label{fD2}}
\end{center}
\end{figure}

\begin{figure}[b]
\begin{center}
\includegraphics[width=7cm]{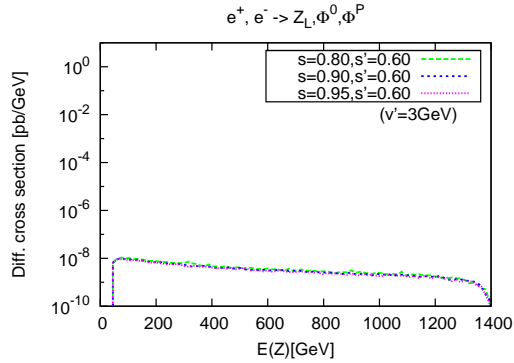}\caption{ Differential cross section versus energy of the $Z_L$ 
boson graphs of the process $e^{+}e^{-}\rightarrow Z_L
\phi^0 \phi^P$ for some selected values of mixing angles when $f=1TeV$ and $v'=3GeV$ at $\sqrt{S}=3TeV$.\label{fD3}}
\end{center}
\end{figure}

For the double production of neutral scalar and pseudoscalar via $e^{+}e^{-}\rightarrow Z_L
\phi^0 \phi^0$, $e^{+}e^{-}\rightarrow Z_L \phi^P \phi^P$ and
$e^{+}e^{-}\rightarrow Z_L \phi^{0} \phi^{P}$ processes, differential cross section versus energy of $Z_L$
 graphs are plotted in figures \ref{fD1}, \ref{fD2} and \ref{fD3} respectively, for $f=1TeV$, and $s/s'=0.80/0.60, 0.90/0.60,0.95/0.60$  at $\sqrt{S}=3TeV$. In 
these calculations VEV of the scalar triplet is taken to be $v'=3GeV$. It is seen from the figures that 
the production rates are not strongly dependent on mixing angles $s/s'$ in the parameter region allowed by electroweak and experimental constraints.

For the production process $e^{+}e^{-}\rightarrow Z_L
\phi^0 \phi^0$, the differential cross section is at the order of $10^{-4}pb/GeV$. The corresponding total cross section is calculated by integrating over $E_Z$ and found to be 
$0.25pb$. At CLIC, the expected luminosity is $100fb^{-1}$, which will result more than few thousands of production events in this channel. At ILC 
the expected center of mass energy is about $0.5-1TeV$, hence this production channel is out of reach due to kinematical limits from high values of scalar mass $M_\phi$.

In the littlest Higgs model, the single production of pseudoscalar $\phi^P$ associated with 
$Z_L$ is not allowed. So the most promising channel for $\phi^P$ production is $e^{+}e^{-}\rightarrow Z_L \phi^P \phi^P$. In this channel, 
the differential cross section is calculated at the order of $10^{-6}pb/GeV$ for all allowed values of mixing angles when $f=1TeV$(Fig. \ref{fD2}). 
The maximum value of cross section for this process is calculated as $3\times10^{-3}pb$ at $\sqrt{S}=3TeV$. For an integrated luminosity of $100fb^{-1}$, up to a few hundreds of $\phi^P$ will be produced within $Z_L$.

For the process $e^{+}e^{-}\rightarrow Z_L \phi^{0} \phi^{P}$, the maximum value of  differential cross section 
is about $10^{-8}pb/GeV$. The corresponding total cross section at $\sqrt{S}=3TeV$ is calculated as 
$10^{-5}pb$. So in this channel the production rate is not promising.


\begin{table}[h]
\caption{The total cross sections in $pb$ for production of neutral
scalars for $f=1TeV$ and at $\sqrt{s}=3TeV$ when $v'=3GeV$ and $v'=10^{-10}GeV$.\label{csecscalars}}
\begin{center}
\begin{tabular}{|c||c|c|}
  \hline  process & $\sigma (pb)[v'=3GeV]$& $\sigma (pb)[v'=10^{10}GeV]$  \\
  \hline
   \hline  ${Z_L\phi^0}$&$10^{-2}$& $10^{-23}$ \\
  \hline  ${Z_L\phi^0\phi^0}$&$0.25$& $2.8 \times10^{-3}$ \\
  \hline  ${Z_L\phi^P\phi^P}$ & $2.8 \times10^{-3}$&$2.7 \times10^{-3}$\\
  \hline  ${Z_L\phi^0\phi^P}$&$1.0\times 10^{-5} $&$1.0\times 10^{-5} $  \\
  \hline \hline
\end{tabular}

\end{center}
\end{table}

A distinguishing feature of neutral scalars in littlest Higgs model is their lepton flavor violating 
decay modes. For lepton flavor violation to be dominant, the VEV of the triplet should be at the order $v'=10^{-10}GeV$. 
For this value, all other decays of the neutral scalar and pseudoscalar are suppressed. In table \ref{csecscalars}, the total cross sections of the 
$Z_L$ associated productions of the neutral scalar and pseudo scalar are given for $s/s'=0.8/0.6$, $f=1TeV$ at $\sqrt{S}=3TeV$ for $v'=3GeV$ and $v'=10^{-10}GeV$.

For the process $e^{+}e^{-}\rightarrow Z_L \phi^0$, at $v'=10^{-10}GeV$, the production cross section is at the order of $10^{-23}pb$. This is due to the explicit dependence of scalar vector vector couplings on the triplet VEV. Thus, for this channel observation of any lepton flavor violation is not possible.

\begin{figure}[h]
\begin{center}
\includegraphics[width=8cm]{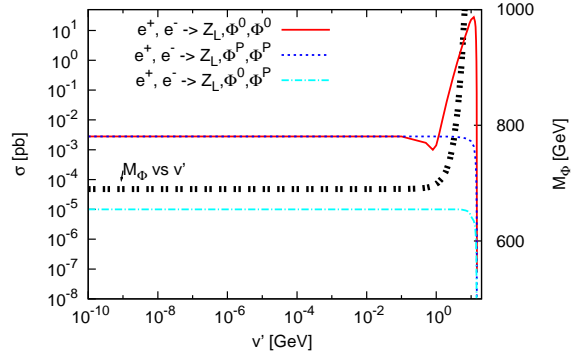}\caption{ Thin lines: Dependence of total cross section on the VEV of the scalar triplet $v'$
 when $f=1TeV$ and $s/s'=0.80/0.60$ at $\sqrt{S}=3TeV$ for $Z_L$ associated pair production of neutral scalar and pseudoscalar. Thick line: Dependence of heavy scalar mass 
$M_\phi$ on $v'$.\label{f5}}
\end{center}
\end{figure}

\begin{figure}[htb]
\begin{center}
\includegraphics[width=7cm]{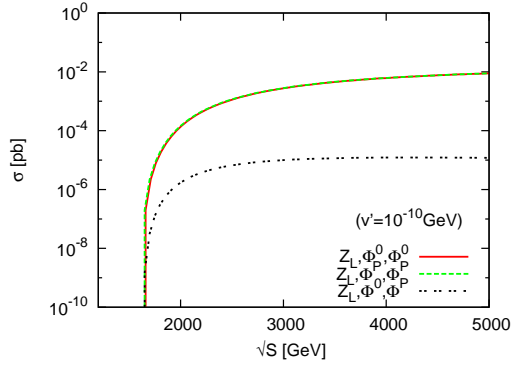}\caption{ Dependence of total cross section on center of mass energy 
 when $f=1TeV$, $v'=10^{-10}$ and $s/s'=0.80/0.60$ for $Z_L$ associated pair production of neutral scalar and pseudoscalar.\label{f5br2}}
\end{center}
\end{figure}

For double production of neutral scalars within $Z_L$, dependence of cross section on $v'$ is plotted in figure \ref{f5}, for $f=1TeV$ at $\sqrt{S}=3GeV$. It is seen that for $v'<0.1GeV$, total cross section is not 
dependent on $v'$. This is due to mass of the heavy scalars, which is steady with respect to variations of $v'$ in this region(Fig. \ref{f5}). In figure \ref{f5br2}, the total cross sections of the processes $e^{+}e^{-}\rightarrow Z_L
\phi^0 \phi^0$, $e^{+}e^{-}\rightarrow Z_L \phi^P \phi^P$ and
$e^{+}e^{-}\rightarrow Z_L \phi^{P} \phi^{0}$ are plotted with respect to center of mass energy, for $f=1TeV$, $s/s'=0.80/0.60$ and $v'=10^{-10}GeV$.
For $v'=10^{-10}$, the total cross section of the processes  $e^{+}e^{-}\rightarrow Z_L
\phi^0 \phi^0$ and $e^{+}e^{-}\rightarrow Z_L \phi^P \phi^P$ are at the order of $10^{-4}pb$ at $\sqrt{S}\sim 2TeV$, and increases 
smoothly to $2.8\times10^{-3}pb$ as center of mass energy approaches to $3TeV$. Since the value of the Yukawa coupling is $Y\sim1$ in this scenario, for an integrated luminosity of $100fb^{-1}$, the number of lepton flavor violating events per year will be close to a thousand. For the process $e^{+}e^{-}\rightarrow Z_L \phi^{P} \phi^{0}$, the total cross section is not 
sufficient to produce lepton flavor violating events. 

\begin{figure}[htb]
\begin{center}
\includegraphics[width=7cm]{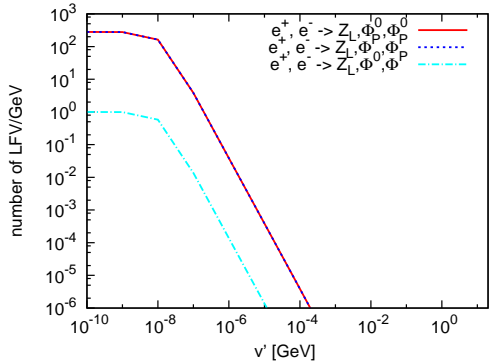}\caption{Dependence of total number of lepton flavour violating final states  on the vacuum expectation value of the scalar triplet $v'$ 
 when $f=1TeV$ and $s/s'=0.80/0.60$ at $\sqrt{S}=3TeV$ for $Z_L$ associated pair production of neutral scalar and pseudoscalar.
\label{f5br}}
\end{center}
\end{figure}

In figure \ref{f5br}, we have plotted the number of 
lepton flavor violating final states with respect to $v'$, for a linear collider with an integrated luminosity of $100fb^{-1}$ at $\sqrt{S}=3TeV$. For these events the collider signature will be \emph{``$Z_L+$missing energy''}. The SM background in this channel is mostly produced via $e^+e^-\to Z_L \nu\bar{\nu}$ processes which has a total cross section of $5pb$. So for this channel, only applying a constraint on the energy of the $Z_L$ boson can improve the signal background ratio. By choosing $Z_L$ bosons carrying the recoil momentum of the scalar pair, i.e. $E_{Z_L}\geq 2 M_{\phi}$, SM contributions suppressed to $3000$ events at $\sqrt{S}\sim 3TeV$. In this case the final state analysis 
will give important results since this signature makes $\phi^0$ and $\phi^P$ indistinguishable but quite different in appearance from their counter 
partners in either SM or two Higgs doublet model.  

In conclusion, the new heavy scalar $\phi^0$ and pseudoscalar $\phi^P$ of the littlest Higgs model will be 
produced  in $e^+e^-$ colliders associated with $Z_L$. The production rates significantly depend on 
the symmetry braking scale parameter $f$ and the vacuum expectation value of the scalar triplet $v'$. 
For $f=1TeV$ and $v'\sim1GeV$ highest production rates are achieved in both channels.
For these parameter set, the productions are quite detectable in the channel $e^{+}e^{-}\rightarrow Z_L \phi^0$ when 
$\sqrt{S}>0.8TeV$ and in the channels $e^{+}e^{-}\rightarrow Z_L
\phi^0 \phi^0$ and $e^{+}e^{-}\rightarrow Z_L \phi^P \phi^P$ when $\sqrt{S}>1.7TeV$. 
However in the channel $e^{+}e^{-}\rightarrow Z_L \phi^0 \phi^P$, there is no significant production rate. For higher values of symmetry breaking scale $f\sim2.5TeV$, the production is achieved only 
in the channel $e^{+}e^{-}\rightarrow Z_L \phi^0$ for $\sqrt{S}\gtrsim1.8TeV$. 
For $v'\sim 1GeV$, the channel $e^+e^-\to Z_L \phi^0$ is the most promising channel for reconstruction of $\phi^0$ from $t\bar{t}$ pairs. The effects of the littlest Higgs model heavy scalar can be observed in $Z_L t \bar{t}$ final states in electron colliders.
For $v'\sim 10^{-10}GeV$ and $f=1TeV$, an interesting and distinguishing feature of the littlest 
Higgs model is on stage. In this case, final decays of the neutral scalar and pseudoscalar are totally lepton flavor violating 
with a collider signature of missing energy accompanied by a SM $Z_L$ boson. For this value of $v'$, the productions in the channel 
 $e^{+}e^{-}\rightarrow Z_L \phi^0$ is not possible, but in the channels $e^{+}e^{-}\rightarrow Z_L
\phi^0 \phi^0$ and $e^{+}e^{-}\rightarrow Z_L \phi^P \phi^P$ the productions of $\phi^0$ and $\phi^P$ are still observable.
Although these channels contain high SM background, the productions and final lepton flavor violating decays of $\phi^0$ and $\phi^P$ 
can still be examined at $e^+e^-$ colliders.

{\small\section{Acknowledgments}
A. \c{C} and M.T.Z. thank to T.M Aliev and A. \c{C} thanks to A. \"Ozpineci, for the guidance and comments.}



\begin{thebibliography}{99}
\bibitem{lh1} N. Arkani-Hamed, A.G. Cohen, E. Katz, and A.E. Nelson, JHEP\textbf{0207}(2002)034, {\tt arXiv:hep-ph/0206021}.
\bibitem{lhmodels1}N. Arkani-Hamed \emph{et al.}, JHEP\textbf{0208}(2002)021, {\tt
arXiv:hep-ph/0206020}.T
\bibitem{lhmodels2}M. Schmaltz,
Nucl.Phys.Proc.Suppl.\textbf{117}(2003),{\tt arXiv:hep-ph/0210415}.
\bibitem{lhmodels3}D.E.Kaplan and M. Schmaltz, JHEP\textbf{0310}(2003)039, {\tt
arXiv:hep-ph/0302049}.
\bibitem{perelstein2ew}J. Hubisz, P. Maeda, A. Noble and M. Perelstein,
JHEP\textbf{01}(2006)135.
\bibitem{B1rizzo} J. L. Hewett, F. J. Petriello and T. G. Rizzo, JHEP\textbf{ 0310} (2003) 062 , {\tt
arXiv:hep-ph/0211218}.
\bibitem{Bdawson} M-C. Chen and S. Dawson, Phys.Rev.\textbf{D70}(2004)015003, {\tt arXiv:hep-ph/0311032}.
\bibitem{Bkilian} W. Killian and J. Reuter, Phys.Rev. \textbf{D70} (2004) 015004, {\tt arXiv:hep-ph/0311095}.
\bibitem{Bdias} A.G. Dias, C.A. de S. Pires P.S. Rodrigues da Silva, Phys.Rev.\textbf{D77}(2008)055001, {\tt arXiv:hep-ph/0711.1154}.
\bibitem{B2csaki} C. Csaki, J. Hubisz, G. D. Kribs, P. Maede and J. Terning, Phys.Rev. \textbf{D68} (2003) 035009 , {\tt arXiv:hep-ph/0303236}.
\bibitem{hakem1} J. A. Conley, J. Hewett and M. P. Le, Phys.Rev.\textbf{D72}(2005)115014, {\tt arXiv:hep-ph/0507198v2}.
\bibitem{perelstein1}M. Perelstein, Prog.Part.Nucl.Phys.\textbf{58}(2007)247-291, {\tt arXiv:hep-ph/0512128}.
\bibitem{thanrev} T. Han, H. E. Logan, B. McElrath and L-T. Wang, Phys.Rev. \textbf{D67}(2003)095004, {\tt arXiv:hep-ph/0301040}.
\bibitem{schmalzrev1} M. Schmaltz and D. Tucker-Smith, Ann.Rev.Nucl.Part.Sci.\textbf{55}(2005)229-270, {\tt arXiv:
hep-ph/0502182}.
\bibitem{atlaswork} G. Azuelos \emph{et al.}, Eur. Phys. J. C. \textbf{3952} (2005) 13.
\bibitem{atlaswork2} F. Ledroit, AIPConf.Proc.\textbf{903}(2007)245-248, {\tt arXiv:hep-ex/0610005}.
\bibitem{LHCreuter}  W.~Kilian, D.~Rainwater and J.~Reuter,
  Phys.\ Rev.\   {\bf D 74}, 095003 (2006) [Erratum-ibid.\  D {\bf 74}, 099905 (2006)], [arXiv:hep-ph/0609119].

\bibitem{A3} C-X. Yue, W. Yang and F. Zhang, Nucl.Phys. \textbf{B716} (2005) 199-214, {\tt
arXiv:hep-ph/0409066}.
\bibitem{cagil1}A. Cagil, M. T. Zeyrek, Phys.Rev. \textbf{D80}055021(2009), {\tt
arXiv:hep-ph/0908.3581}.
\bibitem{cagil2}A. Cagil,  {\tt
arXiv:hep-ph/1010.0102}.
\bibitem{thanlept1} T. Han, H.E. Logan, B. Mukhopadhyaya and R.
Srikanth, Phys.Rev. \textbf{D72} (2005) 053007 , {\tt
arXiv:hep-ph/0505260}.
\bibitem{gaurlept1} S.R. Choudhury, N. Gaur and A. Goyal, Phys.Rev. \textbf{D72} (2005) 097702 , {\tt arXiv:hep-ph/0508146}.
\bibitem{cinlept_L2yue} C-X. Yue and S. Zhao, Eur.Phys.J.\textbf{C50}(2007)897-903, {\tt arXiv:hep-ph/0701017}.
\bibitem{gaurlept2} S.R. Choudhury \emph{et al.}, Phys.Rev.\textbf{D75}(2007)055011, {\tt arXiv:hep-ph/0612327}.
\bibitem{ILC}J. Brau (Ed.) \emph{et al}, By ILC Collaboration, \emph{LC Reference Design Report: ILC Global Design Effort and World Wide
Study.}, FERMILAB-APC, Aug 2007, {\tt arXiv:acc-ph/0712.1950}.
\bibitem{CLIC}R.W. Assmann \emph{et al.}, The CLIC Study Team, \emph{A 3 TeV $e^{+}e^{-}$
linear collider based on CLIC technology}, CERN 2000-008, Geneva,
2000; E. Accomando \emph{et al.}, By CLIC Physics Working Group,
\emph{Physics at the CLIC multi-TeV linear collider.},
CERN-2004-005, {\tt arXiv:hep-ph/0412251}; H. Braun  \emph{et al.},
By CLIC Study Team, \emph{CLIC 2008 parameters}, CERN-OPEN-2008-021,
CLIC-NOTE-764.

\bibitem{neutrinomass} G.L. Fogli \emph{et al}, Phys. Rev. D\textbf{70},
113003(2004);~~M. Tegmark \emph{et al}, SDSS Collabration, Phys.
Rev. D\textbf{69}, 103501(2004).
\bibitem{neutrinomass2}E. W. Otten and C. Weinheimer,~Reports on Progress in Physics, 71 (2008)
086201,~{\tt hep-ex/0909.2104 };~~Z-z Xing, {\tt hep-ph/0907.3014 }.
\bibitem{pdg}K. Nakamura et al. (Particle Data Group), J. Phys. G 37, 075021 (2010).
%
\bibitem{calchep}A. Pukhov \emph{et al.}, CalcHEP/CompHEP Collab., {\tt hep-ph/9908288}
; {\tt hep-ph/0412191}.
%
\end{thebibliography}
\end{document}